\documentclass[11pt]{article}
\usepackage{blois,epsfig}

\def\Journal#1#2#3#4{{#1} {\bf #2}, #3 (#4)}

\def\PLB{{\em Phys. Lett.}  B}
\def\PRL{\em Phys. Rev. Lett.}
\def\PRD{{\em Phys. Rev.} D}

\def\be{\begin{equation}}
\def\ee{\end{equation}}
\def\bea{\begin{eqnarray}}
\def\eea{\end{eqnarray}}

\begin{document}
\vspace*{4cm}
\title{NON--GAUSSIANITY FROM WARM INFLATION IN THE WEAK DISSIPATIVE REGIME}

\author{ S. GUPTA }

\address{Institute of Cosmology and Gravitation, University of Portsmouth,\\ Mercantile House, Hampshire Terrace, Portsmouth,\\Hampshire, England}

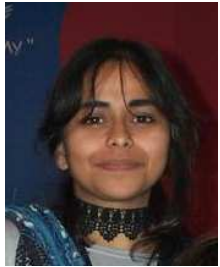
\maketitle\abstracts{
Standard, slow--roll, single--field inflation, as it has been incorporated into standard cosmology, is an over--simplified scenario to which there have been a number 
of suggested physical corrections. The generic prediction for the perturbations generated during slow--roll, single--field inflation, as they appear in the cosmic microwave background (CMB), is a flat, close--to--Gaussian spectrum. We calculate the general solution for a warm inflationary scenario with weak dissipation, reviewing the dissipative dynamics of the two--fluid system, and calculate the bispectrum of the gravitational field fluctuations generated in the case where dissipation of the vacuum potential during inflation is the mechanism for structure formation, but is the sub--dominant effect in the dynamics of the scalar field.}

\section{Introduction}

This work follows on from the work in Gupta {\it et al}~\cite{me}. Gupta {\it et al} calculated the non--Gaussianity expected in the CMB when the temperature fluctuations are generated during warm inflation in the limit of strong dissipation. Our analysis focusses on the dynamics of inflation in the weak--dissipative regime. We give the equations of inflation in the weak dissipative limit, and calculate the non--Gaussianity generated, in order to compare with the predictions from the strong dissipation scenario and standard {\it cool} inflation. 

\section{Warm Inflation}

The standard cosmological model of Hot Big Bang, or Friedmann--Robertson--Walker cosmology with an inflationary phase is highly supported by astrophysical observation (i.e. large scale structure surveys, CMB maps, weak lensing). The challenge remains to make this scenario self--consistent. It has been shown~\cite{bk,bk2} that inflation can occur in the presence of a thermal component to the energy in the Universe.

The subject area of warm inflation covers a range of scenarios, with inflation occurring in the presence of a radiation component. Standard, single--field, slow--roll inflation, with no radiation present during inflation, and perturbations generated in the scalar field density from quantum fluctuations, is one limiting case of warm inflation; the case where the dissipation is set to zero. There have been solutions calculated for the case of warm inflation with strong dissipation~\cite{therm,bt,me}. There is a further limiting possibility of this set of models which is the weak dissipation scenario. There has been much recent interest in this subset of the warm inflation regime. We therefore present the general solutions for polynomial potentials of the evolution of the scalar field for weak dissipative warm inflation.

\section{The Background Dynamics of the 2--Fluid System}\label{subsec:back}

Throughout the wide range of warm inflationary scenarios, the scalar field dominates the dynamics of the system. We begin with these familiar equations
\begin{equation}
\label{eq:rphi}
\rho_\phi=\frac{\dot{\phi}^2}{2}+V(\phi),
\end{equation}
\begin{equation}
\label{eq:pphi}
p_\phi=\frac{\dot{\phi}^2}{2}-V(\phi),
\end{equation}
for the energy density, $\rho_\phi$, and energy pressure, $p_\phi$, of the homogeneous background scalar field, where $V(\phi)$ is the scalar potential and
\begin{equation}
V\gg\frac{\dot{\phi}^2}{2}.
\end{equation}
The warm inflation equation of motion contains a dissipation term which is present during slow--roll.
\begin{equation}
\ddot{\phi}+3H\dot{\phi}+\Gamma\dot{\phi}+V'(\phi)=0\,.
\end{equation}
The prime in the equation above denotes a derivative with respect to $\phi$. Slow roll for this equation of motion now means $|\ddot{\phi}|\ll(3H+\Gamma)\dot{\phi}$.

\subsection{The Strong Dissipative Limit}
In the strong dissipation scenario $\Gamma\gg H$, and in this limit the role of the friction term during inflation is played by the dissipation term.

The stress--energy conservation equation for this strong dissipative case now contains a radiation component and a vacuum energy component:
\begin{equation}
\dot{\rho}_r(t)=-4\rho_r(t) H - \dot{\rho}_\phi(t).
\end{equation}

If there were no dissipation, and therefore no source term, then the radiation component produced would be rapidly red--shifted away, as $\rho_r\sim\exp^{-4Ht}$, due to the exponential expansion of the Universe during inflation. However, with dissipation, radiation is produced continuously from the conversion of scalar field energy. A constant thermal background is maintained~\cite{bk,bk2}.

\subsection{The Weak Dissipative Limit}
The weak--dissipative limit has $H\gg\Gamma$. There is radiation produced via the dissipation which sources the thermal fluctuations in the scalar field. However the Hubble term dominates over the dissipation term, therefore it is the Hubble term which acts as the friction term during slow--roll inflation.

Oliveira \& Joras~\cite{olj} modelled the evolution of perturbations from warm inflation and found that the data as it stands better fits a small value for the dissipation.
 
We derive the energy transfer equations for the limit of weak dissipation.

To model the energy evolution we require a 2--fluid system, of the scalar field, $\phi$, and of radiation, $r$. Labelling the two--fluids with the subscript $\alpha$:
\begin{equation}
\alpha\equiv\{\phi,r\}.
\end{equation}
The stress--energy conservation equations of the scalar field and of the radiation contain an energy transfer term, $Q_\alpha$.
\begin{equation}
\dot{\rho}_\alpha=-3H(\rho_\alpha+P_\alpha)+Q_\alpha
\end{equation}
such that $\sum_\alpha Q_\alpha\equiv 0$.
Looking first at the scalar field evolution, we have
\begin{equation}
\dot{\rho}_\phi=-3H\rho_\phi+Q_\phi.
\end{equation}
We also have, from equations (\ref{eq:rphi}) and (\ref{eq:pphi}), that
\begin{equation}
\frac{\partial\rho}{\partial\phi}\dot{\phi}\sim V'(\phi)\dot{\phi}
\end{equation}
so
\begin{equation}
-3H\dot{\phi}^2+Q_\phi=-3H\dot{\phi}^2-\Gamma\dot{\phi}^2
\end{equation}
and this gives the energy transfer terms
\begin{equation}
Q_\phi=-\Gamma\dot{\phi}^2,
\end{equation}
\begin{equation}
Q_r=\Gamma\dot{\phi}^2.
\end{equation}

\section{Non-Gaussianity of the Perturbations}
The equation of motion of for the full inflaton field in the weak dissipative limit emerges as 
\begin{equation}
\dot{\phi}=\frac{1}{3H}\left[\Delta^2_{\mbox\scriptsize com}\phi(\bf{x},t)-V'(\phi(\bf{x},t))+\eta(\bf{x},t)\right].
\end{equation}
$\eta(\bf{x},t)$ is a Gaussian noise term which models the thermal fluctuations. Our approach to this calculation is stochastic. We expand $\phi(\bf{x},t)$ as a large--scale average and a perturbation:
\begin{equation}
\phi(\bf{x},t)=\phi_0(t)+\delta\phi_1(\bf{x},t)+\delta\phi_2(\bf{x},t),
\end{equation}
where $\delta\phi_2\sim\mathcal{O}(\delta\phi_1)^2$.
A Gaussian field's statistical properties are completely defined by its 1 and 2--point functions. A Gaussian field, with zero mean, will have a bispectrum (the 3--point correlation function in fourier space) predicted to be zero.
\begin{equation}
\langle\Phi(k_1)\Phi(k_2)\Phi(k_3)\rangle=A_{\mbox{\scriptsize inf}}(2\pi)^3\delta^3(k_1+k_2+k_3)\left[P_\Phi(k_1)P_\Phi(k_2)+\mbox{perms}\,\right].
\end{equation}
$\Phi$ represents the peculiar gravitational potential, and $P_\Phi$ represents the power spectrum of perturbations in the gravitational potential.
Even for the cool, slow--roll, single--field inflation, the self--interaction of the inflaton field is known to produce non--zero, but small, non--Gaussian effects. Gangui {\it et al.}~\cite{gang} quantified these effects for various potentials, which give a basis for comparison with supercooled inflation.
We took a general form for the potential
\begin{equation}
V(\phi)=\frac{\lambda}{n!}\,\phi^n\;\;\;, \;\;\;0<\phi<M
\end{equation}
giving, giving for the weak dissipation limit, for $n\neq 2$,
\begin{equation}
\phi(t)=M\left[\frac{M^{n-2}(n-2)}{(n-1)!}\frac{\lambda t}{3H}+1\right]^{-\frac{1}{n-2}},
\end{equation}
and for $n=2$
\begin{equation}
\phi(t)=M\exp\left[-\frac{\lambda}{3H}t\right].
\end{equation}
The relation between the scalar field fluctuations and the gravitational field has, for adiabatic fluctuations, the simple form
\begin{equation}
\Phi(k)=-\frac{3}{5}\frac{H}{\dot{\phi}}\delta\phi(k)\,,
\end{equation}
thus $A_{\mbox{\scriptsize inf}}$ for a weakly dissipative warm inflation regime is
\begin{equation}
A_{\mbox{\scriptsize inf}}^{\mbox{\scriptsize weak}}=-\frac{10}{3}\left(\frac{\dot{\phi}}{H}\right)\left[\frac{1}{H}\ln\left(\frac{k_F}{H}\right)\frac{V'''}{3H}\right]\,.
\end{equation}
$k_F$ is the freeze--out wavevector for each fluctuation mode.
From this, we calculated a value of $A_{\mbox{\scriptsize inf}}^{\mbox{\scriptsize weak}}=7.65\times 10^{-3}$ for the quartic potential. The value of $A_{\mbox{\scriptsize inf}}$ for the strong--dissipative case~\cite{me}, also for the quartic potential, is $A_{\mbox{\scriptsize inf}}^{\mbox{\scriptsize strong}}=7.44\times 10^{-2}$, considerably larger. The corresponding quantity~\cite{gang} $A_{\mbox{\scriptsize inf}}^{\mbox{\scriptsize cool}}=5.56\times 10^{-2}$ is barely distinguishable from the prediction of warm inflation with strong dissipation. 

\section{Conclusions}
The relative non--Gaussianity of the perturbations generated by the energetically motivated scenario of warm inflation with weak--dissipation is different by an order of magnitude from the predictions for strong dissipation and for cool inflation. The difference is not distinguishable in the current CMB data using the bispectrum. Different statistical measures of the CMB non--Gaussianity may yet be able to detect this difference.

\section*{Acknowledgments}
Sujata Gupta is funded by PPARC.

\end{document}